\documentclass[]{aa}
\usepackage[]{psfig}
\usepackage{txfonts}
\begin{document}

\title{On the close optical companion of V1494~Aql: problems with
light curve modelling\thanks{Based on data obtained at the Anglo-Australian
Observatory and Mount Stromlo and Siding Spring Observatories}}

\author{L. L. Kiss\inst{1}\thanks{On leave from University of Szeged, Hungary}
\and B. Cs\'ak\inst{2} \and A. Derekas\inst{1}} 

\institute{School of Physics, University of Sydney 2006, Australia \and
Department of Experimental Physics and Astronomical Observatory,
University of Szeged, Szeged, D\'om t\'er 9., 6720 Hungary}

\titlerunning{The optical companion of V1494~Aql}
\authorrunning{L.L. Kiss et al.}
\offprints{L. L. Kiss,\\
e-mail: {\tt laszlo@physics.usyd.edu.au}}
\date{}

\abstract{New observations of the eclipsing nova V1494~Aql were 
analysed in order to estimate the effect of the optical companion
located 1\farcs4 southwest. $R$ and $I$-band images 
(stellar $FWHM\approx1\farcs2-1\farcs4$) taken close to the primary minimum were 
used to determine relative 
magnitude differences between the nova
and the companion, which were found to be $\Delta R\approx0\fm20$ and $\Delta
I\approx0\fm52$ (in the sense nova {\it minus} companion). After correcting 
$R$-band time-series observations for the secondary
light, the eclipse depth has been found 
to be almost twice as deep. Before modelling any
late eclipsing light curve that is available in the literature, the data must 
be corrected for the light of the companion.
}
 
\maketitle

\section{Introduction}

V1494~Aql (=Nova Aql 1999/2) has been one of the brightest nova outbursts in the
last decade. Shortly after the discovery in December, 1999 (Pereira et al. 1999),
it peaked at $m_{\rm vis}=4$ mag, which has been followed by rapid dimming
($t_3=16$ d, Kiss \& Thomson 2000). The light curve showed a classical transition
phase with quasi-periodic oscillations (see an updated light curve in Kato et al.
2003). Optical spectroscopy (Kiss \& Thomson 2000, Anupama et al. 2001, Arkhipova
et al. 2002, Iijama \& Esenoglu 2003) showed that the nova belonged to the ``Fe
II'' class of Williams (1992) with early expansion velocities up to 2000
km~s$^{-1}$. Spectroscopy in the transition phase suggested the presence  of high
velocity jets, while the  MMRD absolute magnitude (Kiss \& Thomson 2000) has been
combined with spectroscopic reddening determination which placed the star
$1.6\pm0.2$ kpc of the Sun (Iijama \& Esenoglu 2003). Early X-ray observations
were reported by Drake et al. (2003) revealing short-period (2523 s) oscillations
that were interpreted as pulsations of the white dwarf in the nova.
Spectropolarimetric evidence of an asymmetric outburst was presented by
Kawabata et al. (2001).

The photometric history is rich in observations. Kato
et al. (2003) gave an excellent review of the published works, here we just
briefly mention the main results. After the discovery of short-period
modulations of the light curve (Novak et al. 2000), Retter et al. (2000) pointed
out the double-wave nature of the modulation with a peak-to-peak amplitude 
of 0\fm07, 8 months after the outburst. Bos et al (2001) observed a strong
amplitude increase of the modulation and suggested the eclipsing nature of the
system. Barsukova \& Goranskii (2003) refined the orbital period, giving $P_{\rm
orb}=0.1346141(5)$ d. Pavlenko et al. (2003) made a multicolour study of the
star concluding that the light curve shape might be explained by a self
eclipsing accretion column in the magnetic variable accompanied with partial
eclipses of the accretion region by the secondary component. 

We have been following the star since the discovery, mostly monitoring it
spectroscopically (Kiss \& Thomson 2000, Kiss et al. in preparation).
Here we report on new imaging and spectroscopy carried out in 2003, which
revealed the presence of a close optical companion (it has already been
reported by Barsukova \& Goranskii 2003, however, we were not aware of it
before making the observations). Interestingly, none of the recent studies
tried to correct for the photometric effects of the companion. In this paper we
would like to point out the necessity of such a correction.

\section{Observations and data reductions}

\begin{figure*}
\begin{center}
\leavevmode
\psfig{figure=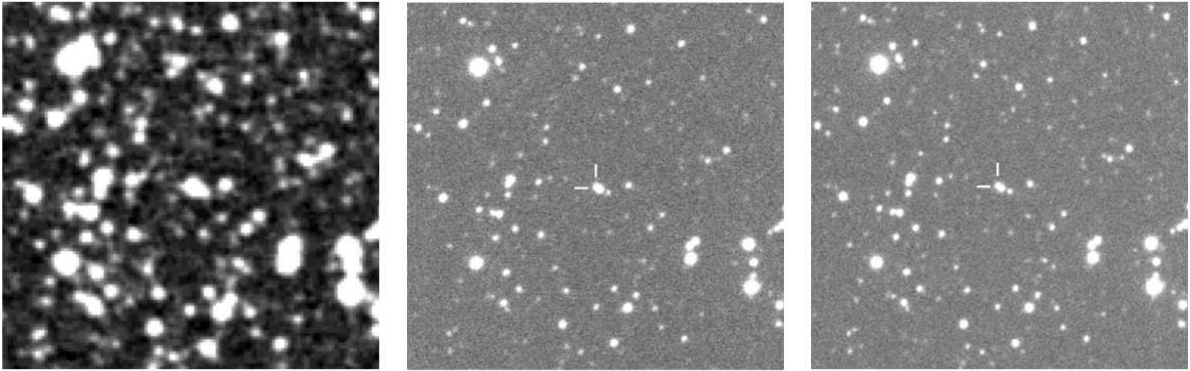,width=16cm}
\caption{$120^{\prime\prime}\times120^{\prime\prime}$ fields centered on
V1494~Aql. Left to right: POSS2 red, $R$-band image at HJD=2~452~930.9568
(stellar $FWHM=1\farcs4$),
$I$-band image at HJD=2~452~930.9526 ($FWHM=1\farcs2$). Dashes point to the nova,
identified through the eclipses.
North is up, east is to the left.}
\label{imagecomp}
\end{center}
\end{figure*}

We discuss three different datasets. In chronological order, they were
taken as follows. On May 10, 2003, between HJD=2~452~770.148--2~452~770.317, four
hours of service observations were carried out using the 3.9m Anglo-Australian
Telescope (AAT) equipped with the RGO spectrograph. We obtained 41
medium-resolution spectra
(exposure time 300 s) covering 5800--7300 \AA, with typical S/N ratios of 50
($\lambda/\Delta \lambda=6000$).
One flux standard and one smooth-spectrum standard star were also observed. The
seeing on that night was about 1\farcs3, as judged from the width of the
standard spectra. The position angle of the 2-arcsec wide slit was constant 
at 200$^\circ$.

Then we made time-series $R$-band observations with the 1.0m telescope of the
Australian National University in Siding Spring on three consecutive
nights in August, 2003 (6-8th), with a total time span of $\sim$12 h. The
detector was the middle 1k$\times$1k subframe of one of the eight 2k$\times$4k
chips of the Wide Field Imager, giving $6\farcm5\times6\farcm5$ field of
view (this corresponds to 0\farcs38/pixel image scale). The
exposure time was 180 s. Since V1494~Aql was a secondary target for bright and
medium to bad seeing nights (the primary program addressed globular clusters),
these time-series data were obtained under 2.5 to 5 arcsec seeings. 

Finally, the last dataset consists of eight CCD images (4 in $R$ and
4 in $I$) taken on Oct. 18, 2003. The instrument was the same as in August, 2003.
The seeing varied between 1\farcs2--1\farcs4,
which has been the best one experienced in a three-weeks long run. The exposure
time was only 30 s in order to get as sharp images as possible. We noticed 
the optical companion of V1494~Aql and that has initiated a careful
re-check of the literature and previous data.

All observations were reduced with IRAF\footnote{IRAF is distributed by the
National Optical Astronomy Observatories, which are operated by the Association
of Universities for Research in Astronomy, Inc., under   cooperative agreement
with the National Science Foundation.} in a standard fashion. Spectroscopic
reductions included bias, flat and sky background corrections. Aperture
extraction and wavelength calibrations were done with the task {\it doslit}
utilizing CuAr spectral lamp exposures taken before and after every ten stellar
exposures. The flux calibration used a spectrum of LTT~7379, a G0-type
spectroscopic standard. Time-series direct images were corrected with bias and
sky-flat frames, while instrumental $R$-band magnitudes were calculated with
simple aperture photometry (the diameter of the aperture was set to 6\farcs5)
in respect to comparison stars located within 1\farcm0. Following 
Barsukova \& Goranskii 2003, the main comparison was GSC 0473--4227. Typical
photometric accuracy was $\pm0\fm01-0\fm02$.

\section{Discussion}

\subsection{Astrometry}

The original purpose of making sharp images on October 18, 2003 was checking 
the presence or absence of a resolvable nova shell. The predicted angular radius
of the shell 3.8 years  
after the outburst is 0\farcs94 (following Kiss \& Thomson 2000 and adopting
the distance revised by Iijama \& Esenoglu 2003). Instead of a shell, we
found two closely separated stars within $\pm$1$^{\prime\prime}$. 
In Fig.\ \ref{imagecomp} we show our best CCD frames compared to
the POSS2 red image (this is the sharpest POSS2 image). It is quite
obvious that the 8-shaped profile is caused by two stars close to the limit of
resolution. The POSS2 image has too low resolution and is too crowded to
recognize this. None of the available POSS, 
POSS2 and 2MASS images of the field shows the nova and its companion resolved,
so that our first conclusion is that we do not have any direct information 
on the progenitor.

The next step was the identification of the eclipsing component.
For this, we used our time-series observations taken 2 months earlier. Even the
best images were made under 2\farcs5 seeing, which disabled earlier recognition
of the pair. Image differences, however, showed that the eclipsing variable
is the eastern component, as has been also concluded by Barsukova \& Goranskii
(2003). 

\begin{figure}
\begin{center}
\leavevmode
\psfig{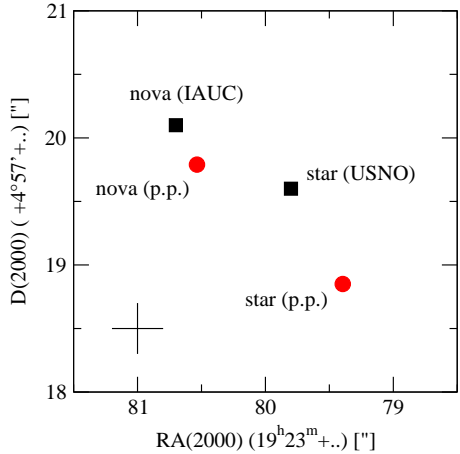}
\caption{A comparison of the new astrometric solution and published
coordinates of the nova and the suspected progenitor (Pereira et al.
1999). The cross in the left corner shows the uncertainty.}
\label{astromet}
\end{center}
\end{figure}

Although the chance of having a nova and an eclipsing binary within
2$^{\prime\prime}$ seems to be very close to zero (the colours reported by
Barsukova \& Goranskii 2003 did not support it either), we made an astrometric
solution of the best $I$-band image and compared the coordinates to the published
values. We chose 21 non-saturated stars within $\pm1^\prime$ of the nova and took
their J2000.0 coordinates from the GSC2.2 catalog. Pixel coordinates were
determined by psf-fitting (with the {\it daophot} package in IRAF). The residuals
of the solution were $\pm$0\farcs2, which we adopted as the astrometric
uncertainty. The resulting coordinates are shown in Table\ \ref{params}. 
We plot them together
with the coordinates of the nova and assumed progenitor (Pereira et al. 1999) in
Fig.\ \ref{astromet}. According to this, the separation and the position angle of
the companion are 1\farcs4 and 230$^\circ$, respectively.  

We have two conclusions based on Fig.\ \ref{astromet}. Firstly, the positional
agreement between the eclipsing star and the nova coordinate measured by D. di
Cicco (Pereira et al. 1999) shows the nova and the eclipsing binary do indeed
coincide. Secondly, the pre-outburst USNO position falls almost exactly
halfway between the nova and the companion. We interpret this as 
an indirect evidence of the similar apparent brightnesses for the progenitor
and the companion.  

\subsection{Photometry}

It is evident from Fig.\ \ref{imagecomp} that the nova has been fainter than
the companion both in $R$ and $I$ (by chance, the
analysed images were taken very close to the primary eclipse, the ephemeris in
Kato et al. 2003 gives $\phi(R)=0.035$ and $\phi(I)=0.002$). This means the
estimated contribution of the companion is much larger than quoted in Barsukova
\& Goranskii (2003) ($\sim$34\% in $R$) and it may present a significant
problem when modelling the light curve. To quantify it, we have measured relative
magnitudes in respect to GSC 0473--4227
($R=12\fm94$, $I=12\fm05$, Barsukova \& Goranskii 2003) and corrected the light
curve accordingly. Neglecting colour-dependent transformation terms (being in
the order of a few hundredth mag for a wide range of colour differences,
Sung \& Bessell 2000), the resulting magnitudes (Table \ \ref{params})
revealed that the nova was fainter by 0\fm20 in $R$ and by 0\fm52 in $I$.
The error budget of these values consists of the individual 
psf-fitting errors calculated by the task {\it allstar} (0\fm05 for the nova and
the companion and 0\fm006 for the comparison), the uncertainty of the
standard magnitudes of the comparison (0\fm02) and the neglected colour 
transformation terms (we assumed 0\fm02). Consequently, the 
final uncertainty is about 0\fm06 (it is marginally better in $I$ thanks to
the better resolution).

\begin{table}
\begin{center}
\caption{Astrometry and photometry of the nova and the companion.}
\label{params}
\begin{tabular}{|lll|}
\hline
    & nova  & companion \\
\hline
RA(2000) &  $19^{\rm h}23^{\rm m}05\fs37$ & $19^{\rm h}23^{\rm m}05\fs29$\\
D(2000) & $04^\circ57^\prime19\fs79$ & $04^\circ57^\prime18\farcs85$ \\
$R$ ($\phi=0.035$) & 16\fm98 & 16\fm78\\
$I$ ($\phi=0.002$) & 16\fm32 & 15\fm80\\
\hline
\end{tabular}
\end{center}
\end{table}

\begin{figure}
\begin{center}
\leavevmode
\psfig{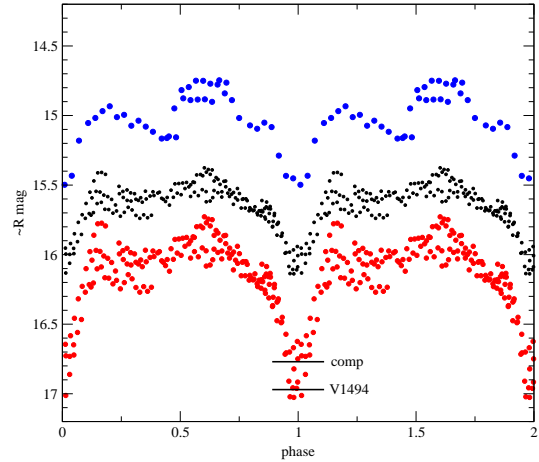}
\caption{Eclipsing light curves. Upper curve: integrated 
magnitudes calculated from the flux-calibrated AAT spectra; middle curve: phase
diagram of all data taken in August, 2003; lower curve: the same after
correcting for the secondary light.}
\label{lc}
\end{center}
\end{figure}

We illustrate the secondary light correction in Fig.\ \ref{lc} (the
companion's magnitude was transformed to flux and was subtracted from the flux
values of the light curve data; the results were transformed back to
magnitudes). The middle curve in Fig.\ \ref{lc} shows the phased $R$-band data
(using the ephemeris in Kato et al. 2003: $BJD_{\rm
min}=2~452~458.3230+0.1346138\times E$), while the corrected curve  is the bottom
one. The eclipse depth (defined as the brightness difference between $\phi=0$
and $\phi=0.15$) has increased by a factor of two from $\sim0\fm6$ to
$\sim1\fm2$, which is obviously not a negligible effect. It is quite surprising
that none of the recent studies applied this correction. On one hand,  
both Barsukova \& Goranskii (2003) and Pavlenko et al. (2003) mentioned the 
presence of the companion but they did not go beyond that. On the other hand, 
despite using unfiltered observations heavily influenced by the red companion, 
Kato et al. (2003) did not even mention the problem. Our conclusion is that
one has to be very careful when analysing observations of V1494~Aql and reliable
modelling requires accurate correction to the light of the companion. 

\begin{figure}
\begin{center}
\leavevmode
\psfig{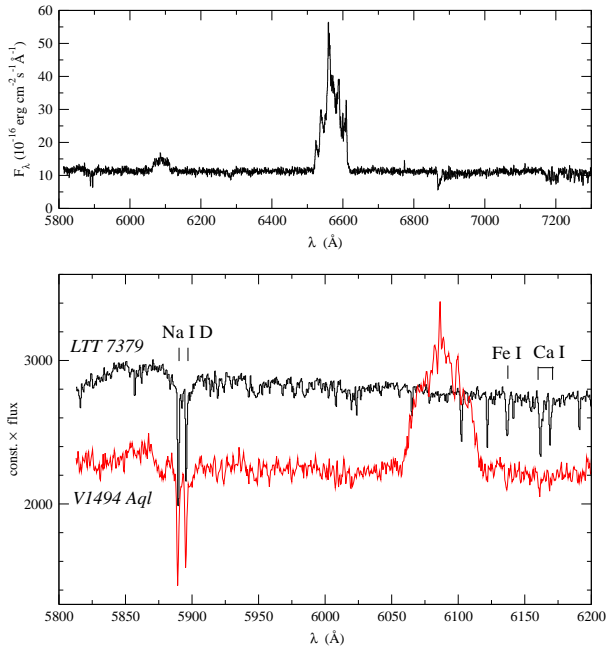}
\caption{Top panel: a sample AAT-spectrum of V1494~Aql. Note the symmetric 
structure of the H$\alpha$ profile. Bottom panel: a close-up view of the bluest
quarter compared to the flux-standard.}
\label{aat}
\end{center}
\end{figure}

\subsection{Spectroscopy}

Further details on the companion's nature are provided by the AAT spectroscopy.
A sample spectrum is shown in the upper panel of Fig.\ \ref{aat}. Beside a flat
continuum, the spectrum is dominated by the complex H$\alpha$ line and an
emission blend at 6080 \AA. Weak absorption lines are present in the blue and the
red third of the spectrum, of which the red one mostly contains atmospheric 
telluric lines. In the blue part the sodium D is the strongest absorption.
The integrated fluxes converted to magnitudes 
draw the same eclipsing light curve as the CCD observations do (see the
top curve in Fig.\ \ref{lc}). After recognizing the problems with the
companion, we checked the original spectrum images and found that nova spectra
were indeed wider than the standard ones (but not wide enough for deblending). 
That means we had the same contamination from the companion as in the case of the
imaging. 

The absence of molecular bands in the spectrum exclude the possibility of a
late-type companion. Furthermore, we could identify a number of common 
absorption lines in the spectra of the nova and the G0-type flux standard.
The most prominent ones are marked in the bottom panel of Fig.\ \ref{aat}. The
resemblance suggests the spectral type of the companion is similar to the flux
standard, most likely between late F and early G. 

Finally, the sodium D line has some interesting properties. 
Iijama \& Esenoglu (2003) detected strong interstellar sodium D 
lines three
months after the outburst. They measured the following equivalent widths:
$EW_{\rm D1}=0.46\pm0.01$ \AA\ and $EW_{\rm D2}=0.55\pm0.01$ \AA. As a
comparison, the blended AAT spectra result in $EW_{\rm D1}=0.50\pm0.02$ \AA\ and 
$EW_{\rm D2}=0.65\pm0.02$ \AA, with possible greater systematic uncertainty  due
to the ambiguous continuum level. The larger equivalent widths are consistent 
with the assumption that we see a sum of stellar and interstellar absorptions. 
Since we do not detect any additional 
line doubling of the
doublet (within $\pm$5 km~s$^{-1}$), either the $\sim$G-type companion has 
similar radial velocity than the cloud in which the interstellar component, 
detected by Iijama \& Esenoglu (2003), originated or the continuum we see 
in Fig.\ \ref{aat} mostly comes from the reddened companion (in which case the
radial velocities of the companion and the cloud still have to agree unless 
the cloud is farther away).

\section{Conclusions}

V1494~Aql is an intriguing eclipsing nova that has initiated recently a number
of  independent photometric and spectroscopic studies. Judged from the
publications, none of them has tried to correct for the effects of the close
companion. With the presented properties of the star, this neglect can be
dangerous and has the potential to lead to unreliable light curve models. The
red and far red spectral regions are especially affected by the companion, which is
a major obstacle to using unfiltered CCD measurements. 

Based on the results presented in this paper, we can summarize our knowledge on
the companion of V1494~Aql. It is located 1\farcs4 SW of the eclipsing nova. In
minimum, the nova is fainter by 0\fm2 in $R$ and 0\fm52 in $I$, while in maximum,
it is brighter by about 0\fm6--0\fm8 in $R$,  depending on the highly variable
light curve shape (see Pavlenko et al. 2003 and the scatter of the phase diagrams
in Fig.\ \ref{lc}). A tentative spectral type has been assigned to the companion
based on medium-resolution spectra with the AAT. 

Adopting $E(B-V)=0\fm6$, $d=1.6$ kpc and assuming the mean corrected $R$ 
magnitude ($\sim16\fm3$) of the nova is close to the quiescent brightness, the
progenitor's absolute magnitude was about $M_{\rm R}=3\fm9$, which leaves
unchanged our conclusion on it (Kiss \& Thomson 2000). However, further
high-resolution imaging, preferably made under sub-arcsecond seeing, is needed 
in order to allow accurate photometric corrections and light curve modelling. 

\begin{acknowledgements}

This work has been supported by the FKFP Grant
0010/2001, OTKA Grants \#F043203 and \#T034615 and the Australian Research
Council. We are grateful to the TACs of the Anglo-Australian Observatory and 
the Mount Stromlo and Siding Spring Observatories for allocating telescope time
used for this study. We also wish to thank the assistance of P\'eter Sz\'ekely
during the time-series photometric observations.
The NASA ADS Abstract Service was used to access data and references.

\end{acknowledgements}

\end{document}